# Silicene on Ag(111): an honeycomb lattice without Dirac bands

S. K. Mahatha[1], P. Moras[2], V. Bellini[2], P. M. Sheverdyaeva[2], C. Struzzi[3], L. Petaccia[3], and C. Carbone[2]

[1]International Center for Theoretical Physics (ICTP), I-34014 Trieste, Italy
[2]Istituto di Struttura della Materia, Consiglio Nazionale delle Ricerche, I-34149 Trieste, Italy
[3]Elettra Sincrotrone Trieste, SS 14, Km 163,5, I-34149 Trieste, Italy

The discovery of (4×4) silicene formation on Ag(111) raised the question on whether silicene maintains its Dirac fermion character, similar to graphene, on a supporting substrate. Previous photoemission studies indicated that the π-band forms Dirac cones near the Fermi energy, while theoretical investigations found it shifted at deeper binding energy. By means of angle-resolved photoemission spectroscopy and density functional theory calculations we show instead that the π-symmetry states lose their local character and the Dirac cone fades out. The formation of an interface state of free-electron-like Ag origin is found to account for spectral features that were theoretically and experimentally attributed to silicene bands of π-character.

The rich physical properties of graphene motivate the study of two-dimensional (2D) materials with honeycomb structure.[1] Silicene, a 2D honeycomb lattice of silicon, is theoretically predicted to be stable as a free-standing monolayer in a low-buckled geometry.[2] The π-symmetry electronic states reach the Fermi energy ($E_F$) at the $K_{Si}$ points of the hexagonal Brillouin zone. They form Dirac cones as in graphene,[2-5] with a comparable, though smaller, group velocity ($0.5 \times 10^6$ m/s in silicene vs. $1 \times 10^6$ m/s in graphene).[6] Similarly to graphene, silicene may display distinct properties that originate from the Dirac cones, with additional characteristics arising from its larger spin-orbit effects.[7,8]

While evidence for free-standing silicene has not been yet experimentally obtained, silicene syntheses by Si deposition on surfaces have been pursued as an alternative method to examine its properties. Most experimental and theoretical studies focus on silicene on Ag(111),[9-19] although honeycomb Si monolayers, with different degree of lattice distortion, have also been reported for Si on ZrB$_2$(0001),[20,21] and Ir(111).[22] Scanning tunneling microscopy (STM) and low energy electron diffraction (LEED) studies identify on Ag(111) honeycomb silicene structures with varying atomic arrangement and significant buckling (0.5-0.8 Å).[10-16] Depending on the growth parameters, Si forms (4×4), ($\sqrt{13} \times \sqrt{13}$)R13.9° and ($2\sqrt{3} \times 2\sqrt{3}$)R30° reconstructed monolayers,[23] mainly differing in their lattice orientation with respect to the substrate axis. The different phases are found to be close in energy, accounting thus for their coexistence under most, or all, growth conditions.[23]

Since the presence of Dirac cones is the characteristic feature that mainly motivates the research on 2D materials, a key, yet unresolved, question is whether silicene preserves them on a supporting substrate. The (4×4) silicene structure on Ag(111) has recently attracted a considerable attention in this respect because it is the silicene phase with the simplest and best characterized atomic structure. An angle-resolved photoemission study of (4×4) silicene on Ag(111) identifies a π-band at the $K_{Si}$ point, which opens a gap at 0.3 eV below $E_F$.[14] This band is interpreted as a branch of a substrate-modified Dirac cone with very large group velocity ($1.3 \times 10^6$ m/s). Another photoemission work finds a similar π-band at the $\overline{M}_{Ag}$ point of the Ag surface, as well as at other equivalent points of the silicene (4×4) reciprocal space.[24] On the contrary, STM measurements under high magnetic field point out the absence of discrete Landau levels as conflicting evidence with Dirac bands close to $E_F$.[19]

The interpretation of the photoemission data[14] has been questioned on the basis of band structure calculations,[19,25-27] which find silicene bands of π-character around 1.1 eV below $E_F$.[19,25] The energy shift of these bands in comparison to free-standing silicene is ascribed to their hybridization with the Ag states. Several theoretical analyses propose that the experimentally reported bands are to be attributed to the Ag substrate, in view of the qualitatively similar dispersion of *sp*-quantum well states in a Ag slab.[19,27] As a matter of fact, however, theoretical analyses until now fail to satisfactorily explain the experimental observations. Contrary to the calculations no π-bands have been experimentally found near 1.1 eV and no theoretical results reproduce the observed silicene-induced band which opens a gap at 0.3 eV binding energy. In consideration of the contrasting experimental and theoretical findings, it appears that the nature of electronic states of silicene on Ag(111) calls to be better established.

We develop a consistent description of the electronic structure of (4×4) silicene on Ag(111) by means of angle-resolved photoemission spectroscopy and density-functional theory (DFT) calculations. We will show that the identification of an interface state, induced by the silicene growth but of free-electron-like Ag character, leads to a substantial revision of former theoretical and experimental descriptions of the system. Our study also highlights how the Si-Ag hybridization differently acts on silicene states of π- and σ-symmetry. The σ-states are moderately affected by the interaction with the substrate, while the π-symmetry states lose their 2D band character. The characteristic Dirac cones of ideal free-standing silicene fade out as a result of the hybridization with the Ag band continuum.

Angle-resolved photoemission studies were performed at the VUV-Photoemission and BaDElPh beamlines of the Elettra synchrotron in Trieste. In both systems the slits of the electron spectrometer were placed at an angle of 45° and 50° with reference to the direction of the p-polarized photon beam, respectively, and photoelectrons were collected within the light scattering plane. Data were acquired with several different photon energies in the 20-150 eV range. Si was deposited by resistive heating of a Si wafer at a rate of 0.01 ML/min on an clean Ag(111) surface maintained at ~240°C, in accordance to the rate used in other works.[10,11,15,16,19] We find in an extended examination of the silicene growth,[23] that the formation of a pure (4×4) silicene phase cannot be achieved on Ag(111). Under the present growth conditions, sharp (4×4) LEED spots were observed, together with a more diffused ($\sqrt{13}\times\sqrt{13}$)R13.9° pattern. DFT calculations were performed using the VASP code[28] with the GGA-PBE exchange-correlation potential[29] and including van der Waals interactions in the semi-empirical method of Grimme.[30]

Fig. 1(a) displays the π- and σ-symmetry band dispersion in free-standing flat (1×1) silicene. It is similar as in graphene, due to the common honeycomb structure and valence occupancy of the two elements. The main difference is the smaller bandwidth of the silicene bands compared to graphene, reflecting the larger atomic distances and reduced orbital overlap. Fig. 1(b) shows the Brillouin zone of the Ag(111) surface, of an ideal unreconstructed (1×1) silicene, and of the (4×4) structure. Due to the additional periodicity, the original (1×1) cell folds into the (4×4) reduced-zone.

Although the superposition of spectral features from different phases may complicate the analysis, it turns out to be possible to separately address the electronic structure of the (4×4) phase by proper consideration of the full geometry of the photoemission data. We will discuss in the following sharp and intense spectral features arising from the (4×4) domains which can be distinguished by their location in the reciprocal space.

Fig. 2 compares photoemission spectra of silicene on Ag(111) with clean Ag(111) data along two high-symmetry directions. We note that all intense spectral features reach their maxima or minima in proximity of the symmetry points of the (1×1) zones of silicene and Ag (or at a multiple wave vectors in the extended zone scheme), with weaker replicas at corresponding points of the (4×4) and ($\sqrt{13}\times\sqrt{13}$)R13.9° cells. This behavior typically arises whenever the potential associated to the supercell represents a small perturbation of the electronic structure of the two individual elements. This observation holds generally in this system for all photon energies and finds a correspondence in the intensity of different order spots in the LEED pattern. Along the $\overline{\Gamma}_{Ag} - \overline{K}_{Ag}$ direction (Fig. 2(a,b)) the bands crossing $E_F$ in silicene can be attributed to the substrate, since they display identical dispersion as in clean Ag, though with different $-k_\parallel$ vs $+k_\parallel$ intensity. No evidence for a gap opening at the silicene $K_{Si}$ point (1.1 Å$^{-1}$) is found in these bands, in contrast with a previous study.[14] We find, however, a silicene-induced band (I) with a maximum at ~0.35±0.05 eV in correspondence of the $\overline{M}_{Ag}$ point, close to the Ag substrate bands (Fig. 2(c,d)). Another study[24] identifies also this band as a π-band, modified by the silicene buckled geometry, in view of the equivalence of the $\overline{M}_{Ag}$ and $K_{Si}$ points when folded in the (4×4) cell (Fig. 1(b)). We will argue in the following, on both experimental and theoretical grounds, that this feature, as well as its weaker replicas in the folded (4×4) zone, originates from an interface state of free-electron-like Ag character.

Other spectral structures appear in Fig. 2(c) near $\Gamma_{Si}$ with maxima at 1.3 eV binding energy. Photoemission data over a more extended binding energy range (not shown) indicate these bands to be the continuation of the deeper lying Si σ-bands with minimum at ~12.5 eV. Other evidence confirming that these structures originate from the σ-band is given by their energy dispersion, that precisely corresponds to the one expected for the top of σ-bands, while it exceeds the one for the π-band by more than a factor two. These structures do not show dispersion with varying photon energy, consistently with their assignment to silicene-related state.

The absence of photon energy dependence of the (I) band indicates that also this band has a 2D character, like surface and interface states. This property would also be consistent with its interpretation as a silicene Dirac cone, modified by the substrate interaction and the buckled structure.[24] However, the Dirac cones are expected to be more prominent at the $K_{Si}$ point, while the

(I) band appears at the $\overline{M}_{Ag}$ point. Clear evidence showing that this band does not represent a modified Dirac cone emerges by a closer examination of its dispersion in the wave-vector space. Fig. 3 presents spectra of silicene/Ag(111) (panels (a,b)) and clean Ag(111) (panels (c,d)) in the vicinity of the $\overline{M}_{Ag}$ point. Similarly to the Ag *sp*-states, the silicene-induced (I) band displays an energy maximum along the $\overline{\Gamma}_{Ag} - \overline{M}_{Ag} - \overline{\Gamma}_{Ag}$ direction and a minimum in the perpendicular $\overline{K}_{Ag} - \overline{M}_{Ag} - \overline{K}_{Ag}$ direction, also recently reported in Ref. [31]. This indicates that the (I) band does not have a conical shape and rules out the existence of a band gap near $E_F$. Fig. 3 also offers a comparison between isoenergy cuts of the silicene/Ag(111) (Fig. 3(e)-(h)) and clean Ag(111) surface (Fig. 3(i)-(l)) in the proximity of the $\overline{M}_{Ag}$ point. The (I) band appears in this representation as an additional contour that splits-off from the Ag *sp*-bands profiles when the bands originating from adjacent Brillouin zones approach the $\overline{M}_{Ag}$ point. The left and right branches of the (I) band merge at $\overline{M}_{Ag}$ for ~0.3 eV binding energy and separate themselves along the orthogonal direction for lower binding energies. Evidently their isoenergy sections do not display the circular shape of Dirac cones at $\overline{M}_{Ag}$, neither above nor below the crossing point. They appear instead to mimic the contour of free-electron-like states of Ag near the $\overline{M}_{Ag}$ point (Fig 3 (i)-(l)), which constitute the dog-bone necks in noble-metal Fermi surfaces. For all tested photon energies we found the (I) state to present the same behavior. The data indicate therefore the formation of a 2D state near the edge of the Ag surface-projected bulk bands[32] that, as we will show through a theoretical analysis, has the character of an interface band. This band has typical free-electron-like dispersion although it is induced by silicene formation. The 2D character of the band, together with the fact that it exists only on silicene-covered Ag(111), excludes that this band simply derives from unperturbed Ag substrate states. No other structure compatible with Dirac cones could be observed at any other wave-vector, despite the used experimental geometry favors the emission of π-bands over that of σ-bands, as shown in previous work on graphene.[33]

In order to shed further light on the silicene band structure we performed a DFT-based slab calculation of a silicene/Ag(111) surface with (4×4) periodicity. We considered a silicene layer adsorbed on top of 15 layers of Ag. The structure was optimized by allowing the top three Ag layers to relax and keeping the bottom twelve layers fixed at the bulk Ag lattice. Fig. 4 reports the projected band structures for (4×4) buckled silicene on Ag(111) (panels (a,b)) and (4×4) buckled unsupported silicene (panel (c)), obtained by peeling it from the Ag substrate. Fig. 4(a) displays the projection of the σ (green) and π (orange) states in the silicene layer, while the *sp* (red) states in the Ag interface layer below silicene are highlighted in Fig. 4(b). The symbol size associated to the Si-π character is enhanced by a factor of two as compared to the Si-σ and Ag-*sp* contribution in panels

(a) and (b), for better visibility. We observe that the Si-σ bands, mostly visible in the region between 2 and 3 eV and at about 1 eV binding energy in panels (a,c), survive the interaction with Ag. The Si-π character, instead, is spread all over the energy spectrum, due to hybridization with the Ag states. The analysis of the local density of states (not shown) supports these findings: Si-π symmetry states lose their resemblance to the ones of a buckled unsupported silicene layer. The main structures of the Si-σ states are instead retained although slightly shifted towards higher binding energies, with respect to the free-standing layer, in good agreement with the structures at $\Gamma_{Si}^{4\times4}$ observed in the experiment. Earlier works attributed these states to Si-π bands,[19,24,25] although no comparative analysis of the relative weight between σ and π characters was provided.

Residual Si-π character is located in bands that closely follow the Ag interface state evidenced in panel (b). This state is mainly localized in the first Ag layer below the silicene and, to a much smaller extent, in the second Ag layer. In order to illustrate the nature of this band Fig. 4(d) displays a sketch of an *sp* free-electron-like bulk Ag band (shown in the inset in the (1×1) reciprocal space) folded into the (4×4) reduced Brillouin zone. The strong similarity between this state (Fig. 4(d)) and the interface state in silicene/Ag(111) (Fig. 4(b)) proves that the latter can be described as a free-electron-like Ag state, localized at the interface by the Si-Ag interaction, rather than a graphene-like M state, as proposed in Ref. [31]. In agreement with the experiments, it shows a saddle-like topology (see the band close to the $\overline{M}_{Ag}$ point in the inset of Fig. 4(d)). The interface state crosses $\Gamma_{Si}^{4\times4}$, where the $\overline{M}_{Ag}$ point folds into, at 0.4 eV binding energy, matching well the (I) band observed in experiments. In addition to showing only minor Si contribution (we estimate at most 20-25% Si character, as compared to Ag), its dispersion in k-space is obviously not compatible with the one expected from a modified Dirac-cone branch.

These findings also provide a straightforward explanation for a Si-induced interface state near 1 eV binding energy at the $K_{Si}$ point, which was discussed in a previous theoretical work.[27] It can be easily realized from Fig. 4(c,d) that such band has no particular significance since it represents just a segment, along the $\overline{\Gamma}_{Ag}$-$\overline{M}_{Ag}$ direction of the whole interface state (I), which reaches its maximum at the $\overline{M}_{Ag}$ point. We also identify in Fig.4(d) some highly dispersive bands, that form around 1 eV binding energy reach 1-2 eV above $E_F$, originating from the branch (on the right side of the inset) which disperses along $\overline{\Gamma}_{Ag} - \overline{K}_{Ag}$. The absence of these bands in Fig. 4(b) testifies that in the vicinity of $\overline{K}_{Ag}$ the interface character is lost, which is compatible with the absence of features induced by Si deposition near $\overline{K}_{Ag}$ in the experiment (Fig. 2(b,c)). The very poor $p_z$ Ag character of the (I) state implies that Si-Ag hybridization takes place not because of direct *sp*-Si/*sp*-Ag interaction but it is mediated by the Ag *d*-orbitals, such as the Si-*p*/Ag-*d* hybridization mechanism observed in transition metal silicides.[34] In addition to showing a dispersion in k-space obviously not compatible with the one expected from a modified Dirac-cone branch, the bands presents only minor Si character, which we estimate to amount to at most 20-25% as compared to the Ag contribution.

The present results lead to significantly revise the description of silicene interactions with the Ag surface. The electronic coupling to the substrate has stronger consequences than assumed in former experimental and theoretical studies. It induces a substantial delocalization of the π-states in the underlying Ag(111) layers with a consequent loss of the π-band properties characteristic of ideal 2D honeycomb materials. It is interesting to note, that the cohesive energy of silicene on the Ag substrate is estimated to be in the range of a weak covalent bonding, ~500 meV, for the 4×4 structure as well as for the other silicene structures.[25] For comparison, this bonding strength is significantly higher than for graphene on any transition metals, including those, as e.g. Ni, Re, Ru, where the hybridization with the *d*-substrate electrons strongly affect the Dirac cones. The larger out-of-plane extension of the Si $3p_z$ states can account for a stronger silicene-substrate bonding with respect to graphene, which retains almost intact π-bands on all noble metals. Since Ag is the only transition metal which does not form intermetallic alloys at Si interfaces, similar or even larger hybridization effects on the silicene π-band, with suppression of their Dirac fermion character, may be conceivably expected to take place on most transition metal substrates.

In conclusion, by angle resolved photoemission spectroscopy and DFT calculations we developed a consistent description of the electronic states of silicene on Ag(111). Experimental and theoretical evidence shows that bands previously attributed on experimental and theoretical ground to π-silicene state derive from an interface Ag state of free-electron-like character. We find that silicene on Ag(111) does not preserve the free-standing electronic properties of a honeycomb lattice. The hybridization with the substrate states significantly modify the silicene bands, to an extent that depends on the state symmetry. While the σ-bands are moderately affected by the substrate, the π-states strongly couple to the Ag band continuum, become delocalized and lose the 2D honeycomb character.

**Figure captions**

Figure 1: (a) Calculated band dispersion of σ (green) and π (orange) states for a free-standing (1×1) silicene layer. (b) Surface Brillouin zones of Ag(111) (green solid line), (1×1) silicene (green dashed line) and (4x4) silicene (red line). Black dots indicate high symmetry points.

Figure 2: Angle-resolved photoemission maps for (a) silicene/Ag(111) and (b) clean Ag(111) along $\overline{\Gamma}_{Ag} - \overline{K}_{Ag}$ measured using photon energy of 126 eV. Angle-resolved photoemission maps for (c) silicene/Ag(111) surface and (d) clean Ag(111) along $\overline{\Gamma}_{Ag} - \overline{M}_{Ag}$ measured using photon energy of 135 eV. Write arrows indicate the position of different $\Gamma_{Si}^{4x4}$ points while yellow boxes indicate the region-of-interest zoomed in Fig. 3. The angle-resolved spectra forming the maps are normalized to equal integral counts.

Figure 3: Angle-resolved photoemission maps of silicene/Ag(111) measured along the (a) $\overline{\Gamma}_{Ag} - \overline{M}_{Ag}$ and (b) $\overline{K}_{Ag} - \overline{M}_{Ag} - \overline{K}_{Ag}$ directions. Angle-resolved photoemission maps of Ag(111) measured along the (a) $\overline{\Gamma}_{Ag} - \overline{M}_{Ag}$ and (b) $\overline{K}_{Ag} - \overline{M}_{Ag} - \overline{K}_{Ag}$ directions. Isoenergy cuts at different binding energy for (e-h) silicene/Ag(111) and (i-l) clean Ag(111) surface. All data were acquired at 31 eV photon energy.

Figure 4: Calculated projected band structure for (a,b) buckled silicene/Ag(111) and (c) buckled unsupported silicene in the (4×4) reduced cell. Green, orange and red symbols indicate states of Si-σ, Si-π and Ag-sp character. (d) Sketch of an Ag-*sp* free-electron-like state folded in the (4×4) cell (the band dispersion in the original (1×1) cell is shown in the inset). The blue region indicates low-dispersive *d* states.

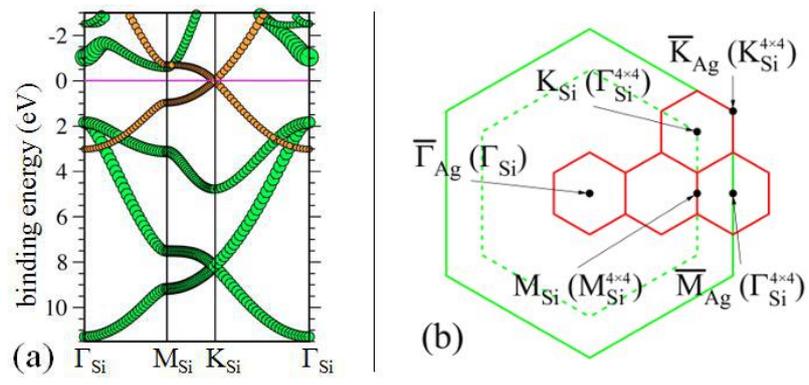

**Figure 1**

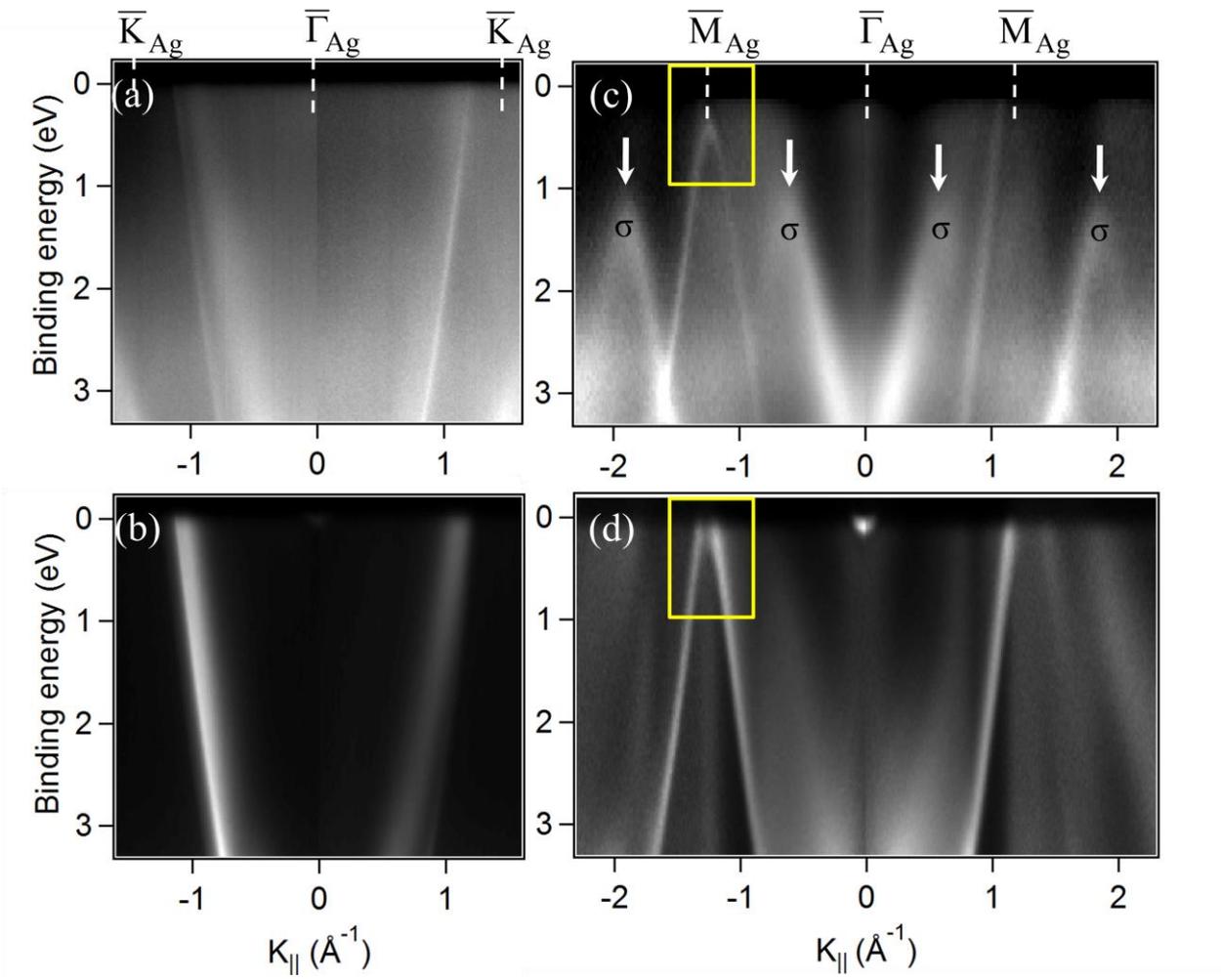

**Figure 2**

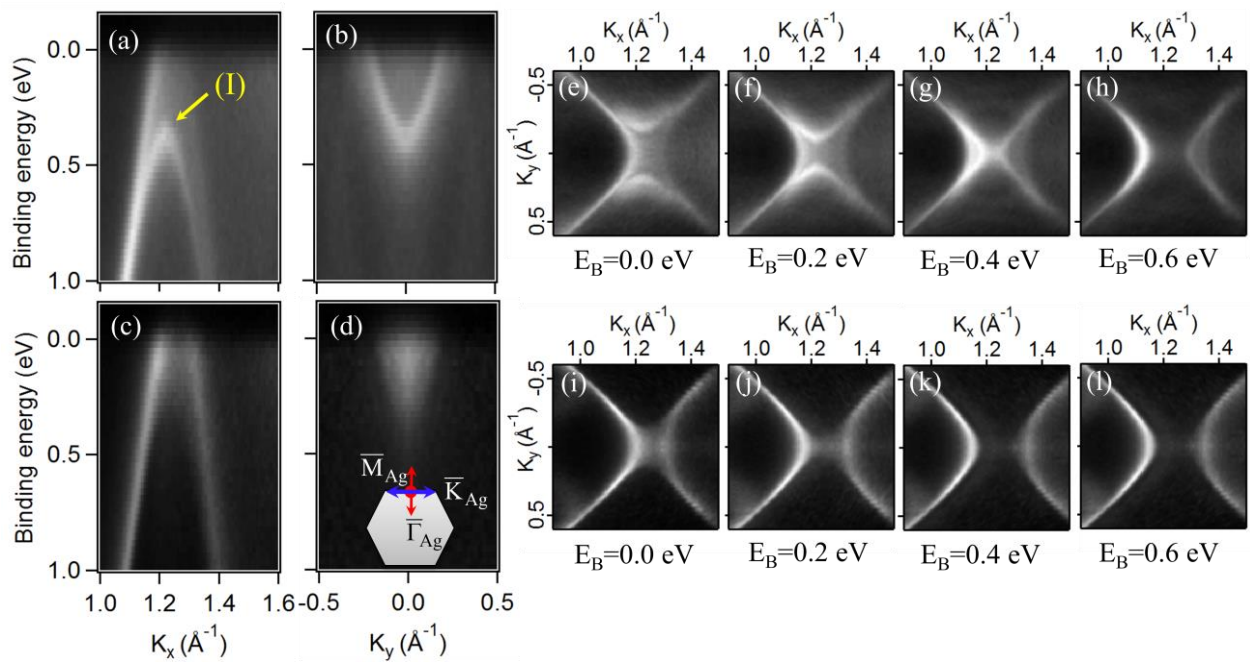

**Figure 3**

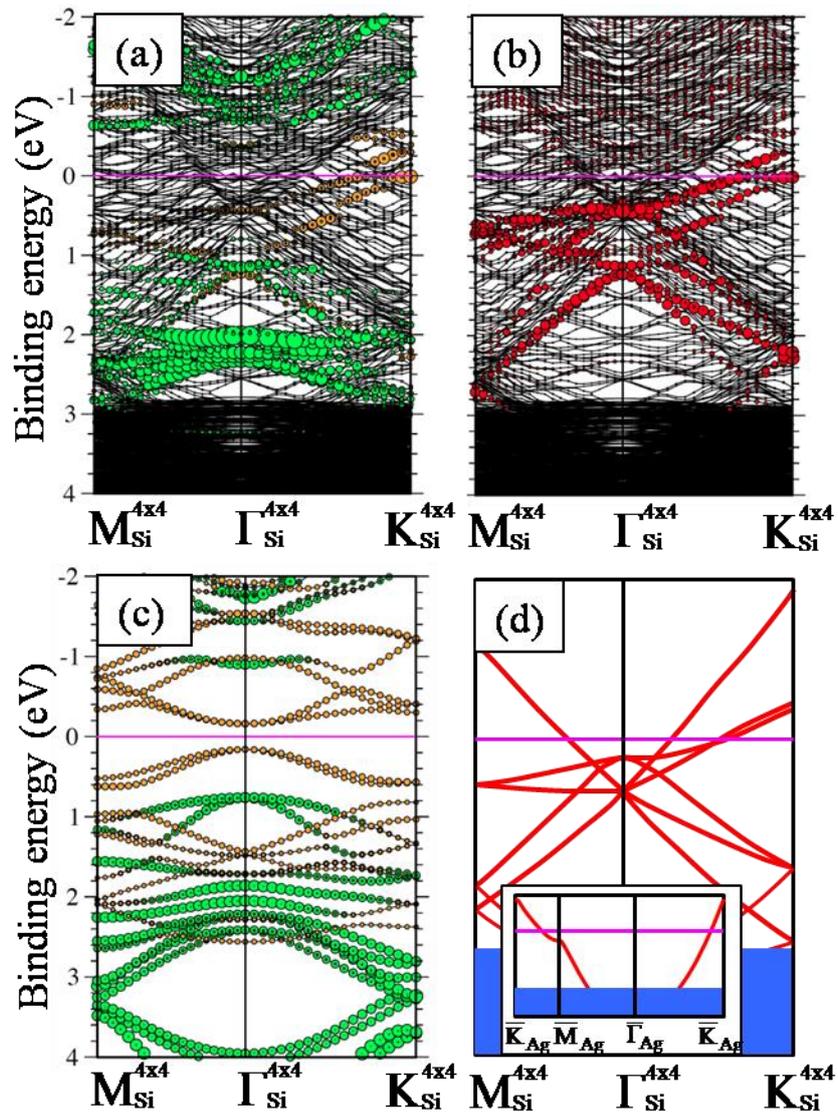

**Figure 4**